\documentclass[a4paper]{aa}
\usepackage{graphicx}
\def \src {1SAX{\thinspace}J1452.8$-$5949}
\def \sax {BeppoSAX}
\def \degmark{^\circ}
\def \nh {N${\rm _H}$}

\def \hcm {\hbox {\ifmmode $ atom cm$^{-2}\else atom cm$^{-2}$\fi}}
\def \arcmin {\hbox{$^\prime$}}
\def \arcsec {\hbox{$^{\prime\prime}$}}
\def \chisq {$\chi ^{2}$}

\def\approxgt{\mathrel{\hbox{\rlap{\lower.55ex \hbox {$\sim$}}
        \kern-.3em \raise.4ex \hbox{$>$}}}}
\def\approxlt{\mathrel{\hbox{\rlap{\lower.55ex \hbox {$\sim$}}
        \kern-.3em \raise.4ex \hbox{$<$}}}}
\begin{document}

   \thesaurus{6(13.25.5;  % X-rays: stars,
               08.09.2;  % stars: individual: src,
               08.14.1;  % stars: neutron,
               08.02.1;  % binaries: close,
               02.01.2)}  % accretion: accretion discs
\title{Discovery of a faint 437~s X-ray pulsar \src}

\author{T. Oosterbroek\inst{1}
 \and M. Orlandini\inst{2}
 \and A.N. Parmar\inst{1} 
 \and L. Angelini\inst{3}\thanks{Universities Space Research Association} 
 \and G.L. Israel\inst{4}
 \and D. Dal Fiume\inst{2}
 \and S.~Mereghetti\inst{5}
 \and A. Santangelo\inst{6}
 \and G. Cusumano\inst{6}
% \and F. Frontera\inst{2}
% \and P. Goldoni\inst{4}
% \and M. Guainazzi\inst{1}
% \and L. Sidoli\inst{4}
% \and L.~Stella\inst{6}
% \and M. Werger\inst{1}
}
 
\institute{Astrophysics Division, Space Science Department of ESA, ESTEC,
           Postbus 299, 2200 AG Noordwijk, The Netherlands
\and 
           Istituto TESRE, CNR, via Gobetti 101, I-40129 Bologna, Italy
\and 
           Laboratory for High Energy Astrophysics, Code 660.2, 
	   NASA/Goddard Space Flight Center, MD 20771, USA
\and
           Osservatorio Astronomico di Roma, Via Frascati 33,
           Monteporzio Catone, I-00040 Roma, Italy
\and
           Istituto di Fisica Cosmica G. Occhialini, CNR, via Bassini 15, 
           I-20133 Milano, Italy
\and
           IFCAI, CNR, Via
           U. La Malfa 153, I-90146 Palermo, Italy
}
\offprints{T. Oosterbroek}
\date{Received 31 August 1999/Accepted 21 October 1999}

\maketitle

\begin{abstract}

A new pulsar, \src, was discovered during a BeppoSAX
galactic plane survey on 1999 July 20 at 
R.A.=$14^{\rm h}\; 52^{\rm m}\; 49\fs3$, 
Dec=$-59\degmark\ 49\arcmin\ 18\arcsec$ (J2000) with a 90\% confidence
uncertainty radius of $50\arcsec$.
Coherent pulsations were detected with a barycentric
period of a $437.4 \pm 1.4$~s. 
The X-ray spectrum can be modeled by a power-law with a
photon index of $1.4 \pm 0.6$ and absorption consistent with the
galactic value in the direction of the source 
($1.9 \times 10^{22}$~atom~cm$^{-2}$).
An Fe K line with a equivalent
width of $\approxgt$1.3 keV may be present in the spectrum. 
The unabsorbed 2--10~keV flux is $9 \times
10^{-13}$~erg~cm$^{-2}$~s$^{-1}$. 
The X-ray properties and lack of an obvious optical counterpart
are consistent with a Be star companion at a distance of 
between approximately 6 and 12~kpc
which implies a luminosity of (4--15)$\times 10^{33}$~erg~s$^{-1}$.

\end{abstract}

\keywords   {X-ray: stars --
             stars: rotation --
             pulsar: general--
             stars: individual: \src }

\section{Introduction}

As well as studying studying the unresolved diffuse galactic emission 
(the galactic ridge X-ray emission), 
one of the main scientific objectives of the BeppoSAX galactic
plane survey is to search for faint X-ray pulsars.
Currently, there are $\sim$80 known accreting X-ray pulsars 
(see Bildsten et al. \cite{bi:97} for a recent review).
Until recently only relatively bright nearby pulsars were
visible due to the limited sensitivity of the available detectors.
This is changing with the discovery by ASCA, BeppoSAX, 
and RXTE of a population of faint 
($\sim$$10^{-11}$--$10^{-12}$~erg~cm$^{-2}$~s$^{-1}$) pulsars
(e.g., Angelini et al. \cite{a:98}; Kinugasa et al. \cite{k:98}; 
Santangelo et al. \cite{s:98}; Sugizaki et al. \cite{s:97};
Torii et al. \cite{t:98}, \cite{t:99}). 
These new pulsars tend to have long periods and it is likely that 
the centrifugal barrier, which inhibits accretion in rapidly rotating
neutron stars at low luminosities
(e.g., Stella et al. \cite{s:86}), is almost always open, allowing them 
to reach very low flux levels. 
It is thus likely that a population of low-luminosity
X-ray binaries is being observed, the emission from which may
contribute significantly to the unresolved galactic component.

\section{Observations}

Results from the Low-Energy Concentrator Spectrometer (LECS;
0.1--10~keV; Parmar et al. \cite{p:97}), and the Medium-Energy Concentrator
Spectrometer (MECS; 1.8--10~keV; Boella et al. \cite{b:97}) 
on-board BeppoSAX are presented. 
The MECS consists of two grazing incidence
telescopes with imaging gas scintillation proportional counters in
their focal planes. The LECS uses an identical concentrator system as
the MECS, but utilizes an ultra-thin entrance window and
a driftless configuration to extend the low-energy response to
0.1~keV. The fields of view (FOV) of the LECS and MECS are 37\arcmin\ 
and 56\arcmin, respectively. 

\begin{figure}
  \centerline{\includegraphics[width=8.5cm]{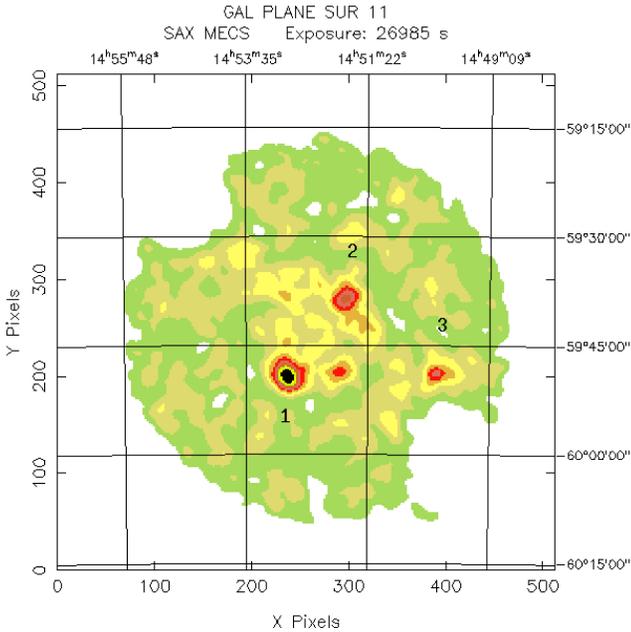}}
  \caption[]{The MECS image (equinox J2000) smoothed
            with a Gaussian filter with a $\sigma$ of 24\arcsec.
            \src\ is the brightest source in the image, located
            just below center (labeled ``1'').
            The two ``cut-outs'' are due to the removal 
            of internal calibration source events}
  \label{fig:image}
\end{figure}

During a systematic survey of part of the galactic plane,
BeppoSAX observed the region of sky around ${\rm l,b = 318\degmark, 
-0.5\degmark}$
between 1999 July 20 11:59 and July 21 02:05~UTC.
Good data were selected from intervals when the elevation angle
above the Earth's limb was $>$$4^{\circ}$ and when the instrument
configurations were nominal, using the SAXDAS 2.0.0 data analysis package.
The exposures in the LECS and MECS are 9.6~ks and 27~ks, respectively. 
The MECS image (Fig.~\ref{fig:image}) reveals 
the presence of at least 3 relatively bright sources. 
Of interest here is a new source located 7\farcm1 off-axis,
with a count rate of $(1.86 \pm 0.07) \times 10^{-2}$~s$^{-1}$ for
both MECS units.
This position is 2\farcm2
from the inner edge of the MECS window support structure. 
The J2000 coordinates, derived from the MECS data, are 
R.A.=$14^{\rm h}\; 52^{\rm m}\; 49\fs3$, 
Dec=$-59\degmark\ 49\arcmin\ 18\arcsec$ 
$({\rm l,b  = 317.645\degmark, -0.463\degmark})$
with a 90\% confidence uncertainty radius of $50\arcsec$.
Although the presence of the window support structure will
bias the centroid of the X-ray counts in the MECS somewhat, the position
determination is limited by the uncertainty in the BeppoSAX 
position reconstruction. We designate the source \src.

\subsection{Timing Analysis}

\begin{figure}
    \centerline{\includegraphics[width=8.5cm]{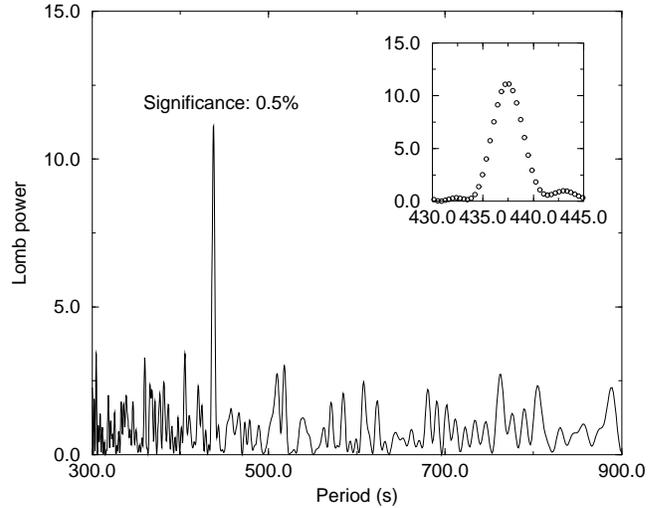}}
    \caption[]{The MECS 1.8--10~keV \src\ Lomb-Scargle periodogram.
               The inset shows the region around the
               probable periodicity in more detail}
    \label{fig:lomb}
\end{figure}

\begin{figure}
    \centerline{\includegraphics[width=8.5cm]{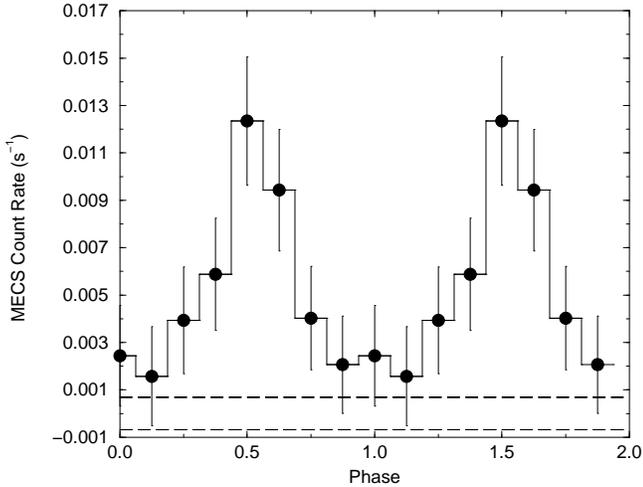}}
    \caption[]{The MECS 1.8--10~keV background subtracted (see text)
    pulse profile. The pulse profile is repeated for clarity. The
    1$\sigma$ uncertainty in the zero-level due to
    uncertainties in the background level is indicated by
    the dashed lines}
    \label{fig:profile}
\end{figure}

A total of 479 MECS events within a radius of 4\farcm7 of \src\ 
were extracted.  The arrival times were corrected to the solar
system barycenter.  Initially, a period search (using the {\sc XRONOS}
routine {\tt efsearch}) in the range 100--1000~s was performed. The strongest
peaks were found at a period of 437~s and at twice this period
(874~s), with \chisq\ values of 42 and 34, respectively, for 8 degrees
of freedom (dof). In order to better asses the significance of
these peaks a Lomb-Scargle periodogram was generated
(Fig.~\ref{fig:lomb}).  A strong peak is detected at 437.1~s. The
chance probability of detecting a peak of this strength or higher in
any of the bins in the Lomb-Scargle periodogram is 0.5\%.
No other strong peaks are evident.  The
period was refined by cross-correlating pulse profiles each obtained
by folding data from 5 consecutive intervals.  This yields a pulse
period of $437.4 \pm 1.4$~s (at 90\% confidence).  The 1.8--10~keV
background subtracted pulse profile (Fig.~\ref{fig:profile}) shows a
peaked modulation with a semi-amplitude (half of the peak-to-peak
modulation divided by the mean) of $74 \pm 24$\%, with an
additional uncertainty due to the uncertainty in the background
subtraction of 10\% (see Fig. \ref{fig:profile}). The pulse profile
does not show any obvious energy dependence.  The
1.8--10~keV lightcurve does not show evidence for variability.
The \chisq\
obtained when fitting a constant value to the lightcurve with 4096~s
bins is 11.9 for 12 dof. The 3$\sigma$ upper limit to any variability
on this timescale is $<$22\%~rms.

\subsection{Spectral Analysis}

LECS and MECS data were extracted centered on the position of \src\ 
using a radius of 4\arcmin\ for both instruments.
Both spectra were rebinned to a minimum of 30 counts per bin, 
and oversampling the full-width at half maximum of the energy
resolution by
at most a factor 3 to allow use of the $\chi^2$ statistic.
Response matrices appropriate to the off-axis source
location in the FOV were used
and spectral fits performed in the 1.0--6~keV (LECS) and
1.8--10~keV (MECS) energy ranges.
Uncertainties are given at 68\% confidence
for one interesting parameter.
Due to the presence of counts from the galactic ridge within the 
source extraction region, background spectra were 
extracted from 3 different source free regions at the
same offset in the FOV as the source and with the same extraction radius. 
Each MECS background
spectrum contains $\sim$250~counts. Fits using these
backgrounds give consistent results, so a mean background was used
in the subsequent analysis. 

\begin{figure}
    \centerline{\includegraphics[height=8.5cm,angle=-90]{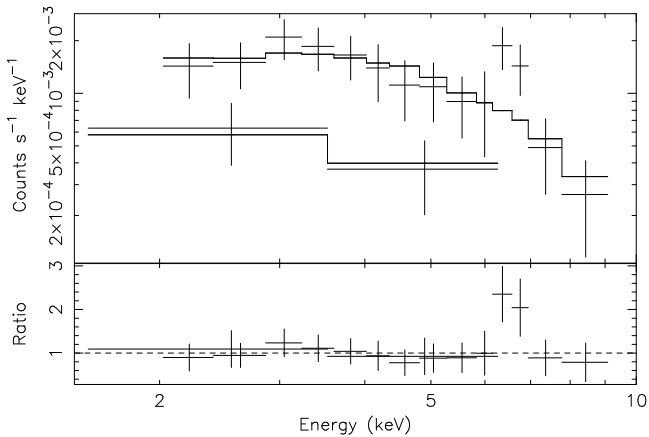}}
    \caption[]{The LECS and MECS source spectra fit with the best-fit
    power-law model. Note the clear deviations from the model at
    $\sim$6.5 keV implying the presence of an Fe emission feature}
    \label{fig:spec}
\end{figure}

The spectrum can be well fit with a power-law model giving a $\chi^2$
of 6.8 for 9 dof, with a photon index, $\alpha$, of $1.4 \pm 0.6$,
when the low-energy absorption, \nh, was fixed at the galactic value
in the direction of the source of $1.9\times 10^{22}$~\hcm\ (Dickey \&
Lockman \cite{d:90}) This fit is shown in Fig.~\ref{fig:spec}. Trials
with \nh\ as a free parameter show that this parameter can only be
constrained to be $<$$6.1 \times 10^{22}$~\hcm.  The
unabsorbed 2--10~keV flux is $7 \times
10^{-13}$~erg~cm~$^{-2}$~s$^{-1}$ (the 2--10 keV flux absorbed by
the galactic value is only
12\% lower).  When a narrow line with an energy of 6.45$\pm ^{0.3}_{0.2}$~keV
and an equivalent width with a 1-$\sigma$ lower limit of 1.3 keV
is added to the model, the \chisq\
reduces to 3.5 for 7 dof and $\alpha$ is $1.9
\pm 0.7$.  Although the line feature appears
quite evident in Fig.~\ref{fig:spec}, an F-test reveals that the
addition of this feature is only significant at a confidence
level of 90\%. A similar line feature is also evident in the
mean MECS background spectrum, but with an intensity $\sim$0.25 of that
above. Thus, unless the background line intensity varies by a factor $\sim$5 
on a spatial scale of 15\arcmin, the probable line feature is
associated with the pulsar. 

Since the source is partly obscured by the MECS entrance window
support structure, a MECS spectrum was extracted with a
2\arcmin\ radius in order to investigate this effect.
The appropriate off-axis response matrix was used. 
The fit parameters so derived are
consistent with those obtained from the spectra extracted
with a 4\arcmin\ radius. The only significant difference is in the
unabsorbed 2--10~keV flux estimate which increases to $9 \times
10^{-13}$~erg~cm$^{-2}$~s$^{-1}$. This is thus the best estimate of the
source intensity.

Although \src\ is in the FOV of the non-imaging High Pressure
Gas Scintillation Proportional Counter (5--120~keV) and Phoswich Detector
System (15--300~keV) instruments, no useful spectral or timing information 
could be extracted from these data. 

\section{Previous Observations}  

We have searched the X-ray catalogs at the HEASARC for any previous
observations of the region of sky containing \src. No previous
detections of a source with a position consistent with \src\ are
reported. The two most sensitive observations of the \src\ region
were by RXTE and ROSAT.

This region of sky was observed twice by RXTE during
scans of the galactic plane (Valinia \& Marshall \cite{v:98}). From
their Fig.\ 1 the upper limit to the strength of any possible
point source near \src\ is $\sim$20~PCA
count~s$^{-1}$. This corresponds to a 2--10 keV flux (using the BeppoSAX
power-law spectral parameters) of $4 \times 10^{-11}$~erg~cm~$^{-2}$~s$^{-1}$,
which is a factor $\sim$40 greater than the intensity observed by
BeppoSAX.

During the ROSAT All-Sky Survey the region of sky containing \src\ was
observed. No source at a position consistent with the new pulsar is
present in the Bright Source Catalogue with an upper limit of
0.05~count~s$^{-1}$ (Voges et al. \cite{v:99}), corresponding to an
unabsorbed flux of $\sim$$2 \times 10^{-11}$~erg~cm~$^{-2}$~s$^{-1}$,
again significantly above the intensity observed here. However, the
ROSAT count rate to flux conversion depends strongly on the amount of
interstellar absorption, which is not well constrained by our
observation.

\section{Distance Estimate}
\label{sect:distance}

\begin{figure}
    \centerline{\includegraphics[height=7.8cm,angle=-90]{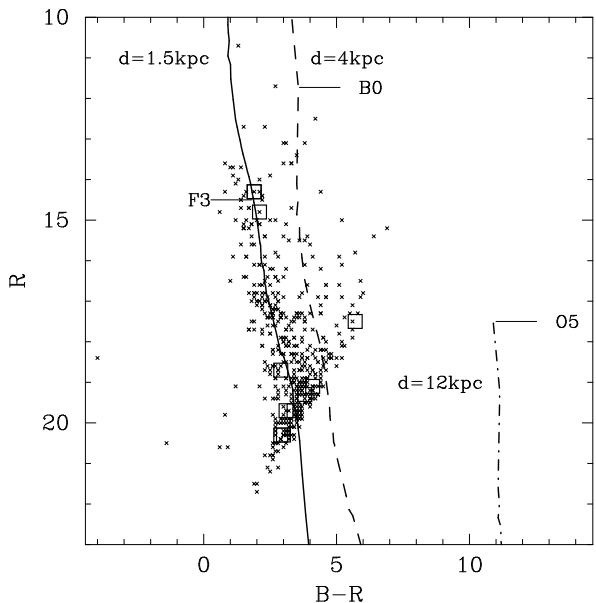}}
    \caption[]{The color-magnitude diagram of stars in the USNO-A
    catalog. Open squares denote stars within the \sax\ error
    region and small crosses denote stars within 5\arcmin. The main
    sequences for 1.5, 4, and 12 kpc are indicated by the continuous, dashed,
    and dot-dashed lines, respectively. The 
    locations of 3 spectral
    types are indicated. Reddening is included assuming linear absorption 
    (which should be a reasonable approximation for d$<$6 kpc),
    with values from Fitzpatrick (\cite{f:99})}
    \label{fig:ccd}
\end{figure}

The line of sight towards \src\ intersects the spiral arms of our
Galaxy at $\sim$1.5, 4, and 12~kpc. We have searched the USNO-A on-line
database for possible optical counterparts.
Fig.\ \ref{fig:ccd} shows the color-magnitude diagram for stars 
located in and around the \src\ error region. 
Assuming that the optical
counterpart is of spectral type O or B (either a main-sequence or a
(super-)giant star, see Sect.~\ref{sect:discussion}), the absence of a
bright blue star in the BeppoSAX error region indicates that the
distance is $\approxgt$6kpc, and probably less than about 12 kpc
(since this is the ``edge'' of the galaxy). This implies a 2--10~keV
luminosity of (4--15)$\times 10^{33}$~erg~s$^{-1}$.

\section{Discussion}
\label{sect:discussion}

Recently, the population of accretion-powered X-ray pulsars has
increased markedly, due to observations with the sensitive detectors
on ROSAT, ASCA, RXTE and BeppoSAX. Bildsten et al. (\cite{bi:97}) list
44 X-ray pulsars and at least 8 others have been recently discovered
(Israel et al. \cite{i:98}; Kinugasa et al. \cite{k:98}; Corbet et
al. \cite{c:98}; Marshall et al. \cite{m:98}, \cite{m:99}; Wijnands \&
van der Klis \cite{w:98}; Hulleman et al. \cite{h:98}; Torii et al.\
\cite{t:99}).  Most X-ray pulsars are in high-mass X-ray binaries
systems (HMXRB) with a few exceptions 
which are in low-mass systems (LMXRB).  \src\ is unlikely to be such a
system since LMXRBs tend to be more luminous and show different types
of variability such as flares, bursting, and transient behavior.  The
large pulse amplitude of \src\ (74$\pm$34\%) is
unusual. While being consistent with the pulse amplitude found in
many HMXRB systems (40--60\%), it may also be substantially
higher. On the basis of the properties reported here,
we cannot exclude the
possibility that the compact object is a magnetized white dwarf.

HMXRB with supergiant companions are 
luminous ($\sim$$10^{35}$--$10^{38}$~erg~s$^{-1}$) systems 
which show marked X-ray intensity
variability. Unless the BeppoSAX observation 
took place during an unusually low state of \src, a supergiant companion seems 
unlikely because of the low luminosity.
More than half of the HMXRB are associated with Be star companions and 
typically show transient behavior. 
The luminosity of Be-star systems during outbursts can change dramatically 
from $10^{33}$ to $10^{38}$~erg~s$^{-1}$, whereas persistent Be system 
such as X Per, or Be systems in quiescence, and e.g.\ the recently
discovered long period pulsars  such as 1WGA J1958.2+3232 (Israel et
al.\ \cite{i:98}) have more modest luminosities 
of $\sim$$10^{33}$--$10^{35}$~erg~s$^{-1}$, consistent with the crude
estimate of the \src\ luminosity in Sect.~\ref{sect:distance}.
Be/X-ray systems display a correlation between their spin and orbital periods 
(Corbet \cite{c:86}; Bildsten et al. \cite{bi:97}) which 
in this case implies an orbital period of  
$\approxgt$200~days for \src.

\begin{acknowledgements}

The BeppoSAX satellite is a joint Italian-Dutch programme.
We thank the staff of the \sax\ SDC for help with 
these observations. 
\end{acknowledgements}

\end{document}